\documentstyle[12pt]{article}

\newcommand\mathC{{\mkern1mu\raise2.2pt\hbox{$\scriptscriptstyle|$}
        {\mkern-7mu\rm C}}}
\newcommand{\mathR}{{\rm I\! R}}                

\renewcommand\[{[\,}                            

\renewcommand\mathR{{\rm I\! R}}

\newcommand\unit{{\rm 1\kern-3.2pt I}}

\begin{document}
\title{General relativity histories theory II: \\
Invariance groups. }
\author{ Ntina Savvidou \thanks{ntina@imperial.ac.uk}\\ {\small Theoretical
Physics Group, The Blackett Laboratory,} \\ {\small Imperial
College,
 SW7 2BZ, London, UK} \\ }

\maketitle

\begin{abstract}

In this paper we show in detail how the histories description of
general relativity carries representations of both the spacetime
diffeomorphisms group and the Dirac algebra of constraints. We
show that the introduction of metric-dependent equivariant
foliations leads to the crucial result that the canonical
constraints are invariant under the action of spacetime
diffeomorphisms. Furthermore, there exists a representation of the
group of generalised spacetime mappings that are functionals of
the four-metric: this is a spacetime analogue of the group
originally defined by Bergmann and Komar in the context of the
canonical formulation of general relativity. Finally, we discuss
the possible directions for the quantization of gravity in
histories theory.
\end{abstract}

\renewcommand {\thesection}{\arabic{section}}
 \renewcommand {\theequation}{\thesection.\arabic{equation}}
\let \ssection = \section
\renewcommand{\section}{\setcounter{equation}{0} \ssection}

\pagebreak

\section{Introduction}

The work presented here is a continuation of \cite{Sav03} in which
we discussed the covariant and the canonical description of
histories general relativity. A key ingredient of \cite{Sav03} was
the introduction of metric-dependent foliations in order to
preserve the spacelike character of a foliation with respect to a
Lorentzian four-metric $g$.

The aim of the present paper is to address the second major issue
of the canonical formalism: namely, the degree to which physical
results depend upon the choice of a Lorentzian foliation. For each
choice of foliation, solutions to the canonical equations of
motion yield different 4-metrics. If different such descriptions
are to be equivalent, the corresponding 4-metrics should be
related by spacetime diffeomorphisms. We must show, therefore,
that the action of the spacetime diffeomorphisms group intertwines
between constructions corresponding to different choices of the
foliation.

To this end, we will discuss the relation between the two major
invariance groups of gravity, namely, the group of spacetime
diffeomorphisms ${\rm Diff}(M)$, and the (canonical) Dirac algebra
of constraints. The key result is that the natural requirement of
the equivalence between descriptions corresponding to different
choices of foliation, can be expressed by a simple mathematical
condition, which we shall call the {\em equivariance
condition\/}\footnote{The term `equivariance' is usually employed
in the following situation. If  two spaces $M$ and $N$ carry
actions $\alpha$ and $\beta$ respectively of a group $G$, then a
map $f: M \rightarrow N$ is {\em equivariant\/} with respect to
these actions, if $\beta(g)f(x) = f(\alpha(g)x))$, for all $x \in
M$, $g \in G$. In the present context the group $G$ is Diff(M).}.

In particular, given a tensor field $A(\cdot, g]$ that is
parameterised by a Lorentzian metric $g$, the possibility arises
that, under a diffeomorphism transformation, the usual induced
transformation of $A$ can be compensated by the additional change
arising from the functional dependence on $g$. Specifically, we
say that the tensor function $g\mapsto A(g)$ is {\em
equivariant\/} if
\begin{equation}
 f^* A(\cdot \,, g] = A(\cdot \,, f^* g]
\end{equation}
for all Lorentzian metrics $g$ and $f\in{\rm Diff}(M)$; here,
$f^*$ denotes the usual pull-back operation on tensor fields.

Of particular importance in what follows is the analogous notion
of an {\em equivariant foliation\/}. Specifically, we say that a
metric-dependent foliation ${\cal
E}(g):\mathR\times\Sigma\rightarrow M $ is equivariant if
\begin{equation}
                {\cal E}(f^*g)=f^{-1}\circ {\cal E}(g)
                \label{Def:EquivariantFoliation}
\end{equation}
for all Lorentzian metrics $g$ and $f\in{\rm Diff}(M)$.

The introduction of equivariant metric-dependent embeddings leads
to a very significant result: the Hamiltonian constraints, the
canonical action functional, and the equations of motion on the
reduced state space are all invariant under the action of the
group of spacetime diffeomorphisms.

In addition to the usual canonical and spacetime
covariance groups, Bergmann and Komar showed that one may
define a group of generalised spacetime diffeomorphisms
that have a functional dependence on the four-metric $g$
\cite{BK}; in what follows we shall denote this by ${\cal
BK}(M)$. We shall show that a representation of this
group also exists in histories theory, and we will
discuss its relation with the other two groups.

\subsection{Relation between the spacetime and the canonical
general relativity description}

The history space for  general relativity is defined as $\Pi^{cov}
= T^{*}{\rm LRiem}(M)$, where ${\rm LRiem}(M)$ is the space of all
Lorentzian four-metrics ${g_{\mu\nu}}$, on a four-dimensional
manifold\footnote{We will assume that $M$ has the topology of
$\mathR \times \Sigma$ for some three-manifold $\Sigma$.} $M$,and
$T^{*}{\rm LRiem}(M)$ is its cotangent bundle. It is equipped with
the symplectic form $ \Omega = \int \!d^4X \, \delta \pi^{\mu\nu}
\wedge \delta g_{\mu\nu}$ where $X \in M$, and $g_{\mu\nu}(X) \in
L {\rm Riem}(M)$, and $\pi^{\mu\nu}(X)$ is the conjugate variable.

The symplectic structure Eq.\ (\ref{omega}) generates the
covariant Poisson brackets algebra,
\begin{eqnarray}
\{g_{\mu\nu}(X)\,, \,g_{\alpha\beta}(X^{\prime})\}&=& 0 \label{covgg}\\
\{\pi^{\mu\nu}(X)\,, \,\pi^{\alpha\beta}(X^{\prime})\} &=& 0
\label{covpipi} \\
\{g_{\mu\nu}(X)\,, \,\pi^{\alpha\beta}(X^{\prime})\} &=&
\delta_{(\mu\nu )}^{\alpha\beta} \,\delta^4 (X, X^{\prime}) ,
\label{covgpi}
\end{eqnarray}
 where we have defined ${{\delta}_{(\mu\nu)}}^{\alpha\beta} :=
\frac{1}{2}({\delta_\mu}^\alpha {\delta_\nu}^\beta +
{\delta_\mu}^\beta {\delta_\nu}^\alpha )$.

For a fixed metric $g$ we can choose a foliation to be spacelike,
in the sense that $t\mapsto h_{ij}(t,\underline{x})$ is a path in
the space of Riemannian metrics on $\Sigma$. However, this
foliation will fail to be spacelike for certain other Lorentzian
metrics on $M$. This is not important at the level of the
classical theory, because we generally only consider four-metrics
are solutions to the equations of motion; however it is a
non-trivial issue in the quantum theory.

In order to address this issue we introduce the space
 of  metric-dependent foliations. For a given Lorentzian metric $g$, we use the
 foliation
${\cal{E}}(g)$ to split $g$ with respect to the Riemannian
three-metric $h_{ij}$, the lapse function $N$ and the shift vector
$N^i$ as
\begin{eqnarray}
 h_{ij}(t,\underline{x}) &:=&
 {\cal{E}}^{\mu}_{,i}(t,\underline{x};g]\,
 {\cal{E}}^{\nu}_{,j}(t,\underline{x};g]\,
 g_{\mu\nu}({\cal{E}}(t,\underline{x};g]) \label{Pullbackh[g]} \\
 N_{i}(t,\underline{x}) &:=&
 {\cal{E}}^{\mu}_{,i}(t,\underline{x};g]\,
 {\dot{\cal{E}}}^{\nu}(t,\underline{x};g]\, g_{\mu\nu}({\cal{E}}
 (t,\underline{x};g])\label{shift[g]} \\
 -N^{2}(t,\underline{x}) &:=&
 {\dot{\cal{E}}}^{\mu}(t,\underline{x};g]\,
 {\dot{\cal{E}}}^{\nu}(t,\underline{x};g]\,
 g_{\mu\nu}({\cal{E}}(t,\underline{x};g]) - N_{i} N^{i}(t,
 \underline{x})
 \label{lapse[g]}
\end{eqnarray}

\bigskip

The symplectic form $\Omega$ can be written in the equivalent
canonical form, with respect to the given foliation as
\begin{eqnarray}
 \Omega &=& \int \!d^4X \, \delta \pi^{\mu\nu} \wedge \delta
 g_{\mu\nu} =  -\int \!d^4X \, \delta \pi_{\mu\nu} \wedge \delta
 g^{\mu\nu}  \label{omega} \\ \nonumber
 &=& \int\!d^3xdt (\delta \tilde{\pi}^{ij}\wedge\delta h_{ij} +
 \delta \tilde{p} \wedge\delta N
 + \delta {\tilde{p}}_{i}\wedge \delta N^{i}), \label{omegacan}
\end{eqnarray}
where
\begin{eqnarray}
\tilde{\pi}^{ij} &:=& \! \! K(t,\underline{x})
(\bar{E}\pi)_{\mu\nu}
 h^{ik} \, h^{jl} \,{\cal{E}}^{\mu}_{,k} \,{\cal{E}}^{\nu}_{,l}
  \label{gpijtilde}  \\
\tilde{p}\!\!&:=& \vspace*{2cm}- K(t,
\underline{x})\frac{2}{N}{(\bar{E}\pi)}^{\mu\nu} n_{\mu}
n_{\nu} \\
\tilde{p}_i \!\!&:=& \vspace*{2cm} - K(t, \underline{x})
\,{(\bar{E}\pi)}_{\mu\nu} (n^{\mu} \dot{{\cal E}}^{\nu}_i +
n^{\nu} \dot{{\cal E}}^{\mu}_i)
\end{eqnarray}
Here $K(t,x)$ is the determinant of the transformation from the
$X$ to the $(t,x)$ variables,
\begin{eqnarray}
K(t, \underline{x}) = \frac{N(t, \underline{x})
\sqrt{\tilde{h}}(t, \underline{x})}{\sqrt{-g}({\cal E}(t,
\underline{x}))}.
\end{eqnarray}
and $\tilde{h}$ is the determinant of the matrix $h_{ij}$.

The kernel $\bar{E}$ stands for $\bar{E}^{\mu \nu}_{\rho
\sigma}(X,X')$, which depends on the chosen foliation; its
explicit form is given in \cite{Sav03}. We should note that for a
foliation with no metric dependence $\bar{E}$ is the unit
operator.

It can be shown that the histories space $\Pi^{cov}$ is equivalent
to the canonical histories space $\Pi^{can} = {\times}_{t}
(T^{*}{\rm Riem}({\Sigma}_{t}) \times T^{*}Vec({\Sigma}_{t})
\times T^{*}C^{\infty}({\Sigma}_{t}))$, where ${\rm
Riem}({\Sigma}_{t})$ is the space of all Riemannian three-metrics
on the surface ${\Sigma}_{t}$, $Vec({\Sigma}_{t})$ is the space of
all vector fields on ${\Sigma}_{t}$, and
$C^{\infty}({\Sigma}_{t})$ is the space of all smooth scalar
functions on ${\Sigma}_{t}$.

The plan of the paper is as follows. In  Section 2.\ we discuss
the status of the symmetries of the theory,  in both the covariant
and the canonical description.  We then discuss the condition for
the physical equivalence of canonical quantities related by
different choices of foliation, and we write its mathematical
expression: the equivariance condition. Furthermore, we elaborate
on the relation between the three symmetry groups.

In Section 3.\ we derive the important result that the histories
reduced state space $\Pi^{red}$ is invariant under spacetime
diffeomorphisms. We conclude with some comments on the possible
quantisation  of gravity within the histories scheme.

\section{Invariance Groups}

The core of this work is the study of invariance groups of the
histories description of general relativity. Starting from the
important result of the co-existence of representations of both
the diffeomorphism group ${\rm Diff}(M)$ and the Dirac algebra of
constraints, we first study the way that invariance
transformations appear in the covariant and in the canonical
description of the histories general relativity. Next, we examine
the relations between the invariance groups, and their special
role in defining general relativity `observables'.

We should mention here that the canonical description of
histories general relativity is not an analogue of the
standard Lagrangian formulation. We will relate the
respective invariance groups of the spacetime {\em and\/}
the canonical descriptions, and we will construct the
spacetime analogue of canonical variables. However, we do
not write the Lagrangian action functional, and in this
sense we do not directly relate it to its canonical
analogue. The direct connection would entail the explicit
relation between Lagrangian and Hamiltonian quantities,
through a histories analogue of the Legendre
transformation.

In what follows, the existence of a representation of the
${\rm Diff}(M)$ group will be shown to be of major
significance for identifying canonical general relativity
observables.

\subsection{Invariance transformations of the covariant description}

The dynamical laws of general relativity are invariant
under spacetime diffeomorphisms (the group ${\rm
Diff}(M)$). However, general relativity is characterised
by a much larger symmetry group also; these are
transformations that are not just point mappings in a
given four-dimensional spacetime, but rather
diffeomorphisms parameterised by the four-metric $g$.

\subsubsection{${\rm Diff}(M)$ active transformations}

The active interpretation of diffeomorphisms
transformations highlights one of the main consequences
of general covariance: spacetime points have no
ontological significance \cite{hole}. A related feature
is that solutions to the field equations that are related
by spacetime diffeomorphisms are regarded as being
physically equivalent.

In \cite{SavGR01, Sav03} we have defined the generator of the
diffeomorphisms group ${\rm Diff}(M)$ to be the generalised
`Liouville' function $V_W$ associated with any vector field $W$ on
$M$ as
\begin{equation}
V_W:=\int \!d^4X \,\pi^{\mu\nu}(X)\,{\cal L}_W g_{\mu\nu}(X)
\label{Vw}
\end{equation}
where ${\cal L}_W$ denotes the Lie derivative with respect to $W$.

These functions $V_W$, defined for any vector field $W$,
satisfy the Lie algebra of the spacetime diffeomorphism
group ${\rm Diff}(M)$
\begin{equation}
\{\, V_{W_1}\,, V_{W_2}\,\} = V_{ [ W_1 , W_2 ]},
\end{equation}
where $[ W_1 , W_2 ]$ is the Lie bracket between vector fields
$W_1$ and $W_2$ on the manifold $M$.

The action of $V_{W}$ on the basic variables of the theory is
expressed by infinitesimal diffeomorphisms,
\begin{eqnarray}
 \{\, g_{\mu\nu}(X)\,, V_W \,\} &=& {\cal L}_W g_{\mu\nu}(X)
 \\[2pt]
 \{\, \pi^{\mu\nu}(X)\,, V_W \,\} &=& {\cal L}_W \pi^{\mu\nu}(X).
\end{eqnarray}

\subsubsection{The histories representation of the Bergmann-Komar
group ${\cal BK}(M)$}

In this section we will show that the covariant formalism
of histories canonical general relativity also carries a
representation of the group of spacetime mappings that
are {\em functionals} of the four-metric $g$; this group
was initially introduced by Bergmann and Komar in
\cite{BK}. We shall start by presenting a brief summary
of their construction.

\paragraph*{The Bergmann-Komar group ${\cal BK}(M)$.}

Originally motivated by the need to identify the
observables in general relativistic theories, Bergmann
and Komar studied the relation between the three major
invariance groups of the theory, namely the group of
spacetime diffeomorphisms, ${\rm Diff}(M)$, the group of
metric-dependent spacetime diffeomorphisms,
$\cal{BK}(M)$, and the Dirac algebra of constraints.

The most general type of spacetime transformations are of
the form
\begin{equation}
 X^{\prime} = f(X; g]
\end{equation}
which are diffeomorphisms that are functionals of the
spacetime metric $g$. The standard spacetime point
mappings in ${\rm Diff}(M)$ are a special case of these
transformations.

Bergmann and Komar state that, in order to identify
general relativity observables, one must find functionals
of the field variables that are invariant under these
general  spacetime mappings. Hence, the three distinct
invariance groups provide three distinct criteria for
selecting the observables (`gauge-invariant' variables).

An infinitesimal transformation, generated by a vector field
$\xi$, i.e., $\delta x^{\mu} = x^{\prime \mu} - x^{\mu} =
{\xi}^{\mu}$, transforms the four-metric $g_{\mu\nu}$, so that
\begin{equation}
 \delta g_{\mu\nu} = g^{\prime}_{\mu\nu} - g_{\mu\nu} = - (\nabla_{\mu}
 \xi_{\nu} + \nabla_{\nu} \xi_{\mu}) = - {\cal L}_{\xi}
 g_{\mu\nu}.
\end{equation}

In general, the vector field $\xi$ can be an arbitrary
functional of the metric, i.e.,
\begin{equation}
 \xi^{\rho} = \xi^{\rho}(X, g], \label{ksi(x,g)}
\end{equation}
where, for the special case of the normal spacetime
diffeomorphisms $\xi^{\rho}$ is a function of $X$ only:
\begin{equation}
 \xi^{\rho} = \xi^{\rho}(X) \label{ksi(x)}.
\end{equation}

Let us now consider the commutator of two consecutive
transformations, namely the vector field $\xi_c$
corresponding to $\delta g_{\mu\nu} = \delta_1 \delta_2
g_{\mu\nu} - \delta_2 \delta_1 g_{\mu \nu}$. For the case
of (\ref{ksi(x)}), it is merely the Lie bracket of two
vector fields, for example $\xi_1$ and $\xi_2$,
\begin{equation}
 \xi_c^{\mu} = {\xi_1^{\mu}}_{, \, \rho}\, \xi_2^{\rho} -
 {\xi_2^{\mu}}_{,\, \rho} \, \xi_1^{\rho} = -[ \xi_1 ,
 \xi_2]^{\mu}. \label{gendif}
\end{equation}

In the more general case (\ref{ksi(x,g)}), the commutator is more
complicated,
\begin{eqnarray}
 \xi_c^{\mu}\!(\!X\!)\!\!\!\!&=& \!\!\!\!{\xi_1^{\mu}}\!_{, \, \rho}\, \xi_2^{\rho} -
 {\xi_2^{\mu}}\!_{,\, \rho} \, \xi_1^{\rho} \nonumber  \\
 &=& \!\!\!\!\! -[ \xi_1 , \xi_2]^{\mu}\!\! - \!\!\!\int\!\! d^4\!X^{\prime}
 \{ \frac{\delta
 \xi_1^{\mu}\!(X)}{\delta g_{\alpha \beta}\!(\!X^{\prime}\!)}\! {\cal{L}}\!_{\xi_2}
 g_{\alpha\beta}\!(\!X^{\prime}\!) - \frac{\delta \xi_2^{\mu}\!(X)}{\delta g_{\alpha
\beta} \!(\!X^{\prime}\!)} \!
{\cal{L}}\!_{\xi_1}g_{\alpha\beta}\!(\!X^{\prime}\!) \}.
\label{genBK}
\end{eqnarray}

For the case of $\xi^{\rho} = \xi^{\rho}(X)$, we obtain
just the expression (\ref{gendif}). Hence, it is an
obvious result that the ${\rm Diff}(M)$ group is a
subgroup of the enlarged diffeomorphisms group
${\cal{BK}}(M)$.

Next, Bergmann and Komar claim that the Dirac algebra of
constraints is a subalgebra of the algebra of
${\cal{BK}}(M)$. In what follows we show the existence of
a representation of the ${\cal{BK}}(M)$ on the history
space $\Pi^{cov}$, and we will study the explicit
relation between the two algebras.

The histories space $\Pi^{cov}$ carries a symplectic
representation of the enlarged (metric-dependent) diffeomorphisms
group ${\cal{BK}}(M)$. We write the histories generator of the
${\cal{BK}}(M)$ group,
\begin{equation}
 U_W :=\int \!d^4X \,\pi^{\mu\nu}(X)\,{\cal L}_W
 g_{\mu\nu}(X),
\label{Vbkw}
\end{equation}
where now, the vector field $W$ is a functional of the
four-metric $g$.

We first write the commutators of the generator $U_W$ with the
field variables $g_{\mu\nu}$ and $\pi^{\mu\nu}$,
\begin{eqnarray}
 \{ U_W , g_{\mu\nu} \}\! &=& \! - {\cal L}_W
 g_{\mu\nu}   \label{Vbkw,g} \\
 \{ U_W , \pi^{\mu\nu} \}\! &=& \! - {\cal L}_W \pi^{\mu\nu}+
 2\!\int\!\! d^4\!X^{\prime} \frac{\delta {\xi}^{\rho}(X^{\prime})}
 {\delta g_{\mu \nu}(X)} \nabla^{\prime}_{\sigma}
 {\pi}^{\sigma\tau}(X^{\prime}) g_{\rho \tau} (X^{\prime})  \label{Vbkw,pi}.
\end{eqnarray}

It is straightforward to show that the generators $U_W$
satisfy the Lie algebra of the ${\cal{BK}}(M)$ group,
\begin{equation}
 \{ U_{W_1} , U_{W_2} \} = U_{W_3},
\end{equation}
where the vector field ${W_3}$ is given by the expression
(\ref{genBK}) with ${W_3}$ being the vector field
${\xi}_c^{\mu}$ of this expression.

In a later section we will examine the relation between
the histories representation of the Bergmann-Komar group,
the spacetime diffeomorphism group, and the Dirac algebra
of constraints.

\subsection{Symmetries of the canonical description}

In \cite{Sav03} we showed that the Dirac algebra of constraints
also appears on the history space. Its generators are\footnote{For
reasons of simplicity we do not introduce the density $\alpha(t)$
of \cite{Sav03}. In what follows, we use the densities
$\tilde{\pi}^{ij}, \tilde{p}_i, \tilde{p}$ instead of the
quantities $\pi^{ij} = \alpha(t)\tilde{\pi}^{ij}, p_i = \alpha(t)
\tilde{p}_i, p = \alpha(t) \tilde{p}$, which are scalars with
respect to time. This means that $\alpha(t)$ does not appear in
the expressions for the constraints. }
\begin{eqnarray}
{\cal H}_\perp(t,\underline x)&:=&\kappa^2
\tilde{h}^{-1/2}(t,\underline{x})(\tilde{\pi}^{ij}(t,\underline{x})
\tilde{\pi}_{ij}(t,\underline{x})
- \frac{1}{2} (\tilde{\pi}_i{}^i)^2(t,\underline{x})) - \nonumber \\
&&\hspace{1cm}
\kappa^{-2}\tilde{h}^{1/2}(t,\underline{x})
R(t,\underline{x}) \label{HperpHis}\\[3pt]
\hspace{-1cm} {\cal{H}}^i(t,\underline x)&:=& - 2 {\nabla}_{\!\!j}
\tilde{\pi}^{ij}(t,\underline x).
\end{eqnarray}

The smeared form of the super-hamiltonian
${\cal{H}}_{\bot} (t,\underline{x})$ and the
super-momentum ${\cal{H}}_i (t,\underline{x})$ history
quantities are defined using as their smearing functions
a scalar function $L$ on spacetime $M$, and a vector
field $\vec L$ that is spacelike in the sense that
\begin{eqnarray}
L^{\mu}(X;g) n_{\mu}(X;g] = 0
\end{eqnarray}
where the one-form $n_{\mu}(X;g]$ is the unit normal to
the leaves of the foliation. The corresponding covariant
expression for the constraints, which is necessary for
relating their action to that of the diffeomorphism
group, are
\begin{eqnarray}
{\cal{H}}[ \vec{L} ] = \int d^4X (\bar{E} \pi)^{\mu \nu} {\cal
L}_L g_{\mu \nu} + 2 \int d^4X (\bar{E} \pi)^{\mu \nu} n_{\mu}
n^{\rho}
{\cal L}_L g_{\rho \nu} \\
{\cal H}_\perp[L] = \int d^4 X \left[\kappa^2
\frac{N}{\sqrt{-g}} \,\frac{1}{2}G_{\mu \nu \rho \sigma}
(\bar{E}\pi)^{\mu \nu}\! (\bar{E}\pi)^{\rho \sigma} -
\kappa\!^{-2} \frac{\sqrt{-g}}{N} {}^3R(h)\right].
\end{eqnarray}

 The tensor $G_{\mu \nu \rho \sigma}$ is the
"Dewitt metric"
\begin{eqnarray}
G_{\mu \nu \rho \sigma} = h_{\mu \rho} h_{\nu \sigma} + h_{\mu
\sigma} h_{\nu \rho} - h_{\mu \nu} h_{\rho \sigma},
\end{eqnarray}
where $h_{\mu \nu} := g_{\mu \nu} + n_{\mu} n_{\nu}$.

Furthermore, we add the primary constraints $\Phi(k)=0$,
where
\begin{equation}
\Phi(k) := \int d^4X (\bar{E} \pi)^{\mu \nu} n_{\mu}(X;g]
\,k_{\nu}(X),
\end{equation}
in terms of a smearing one-form $k_{\mu}$.

It is interesting to note that the supermomentum
constraint
 reads ${\cal{H}}[ \vec{L} ] =
 Q[\vec{L}] + 2 \Phi(n \cdot
{\cal L}_L g)$, where
\begin{eqnarray}
 Q[\vec{L}] :=  \int d^4X (\bar{E} \pi)^{\mu \nu} {\cal
L}_L g_{\mu \nu} \,.
\end{eqnarray}
This expression will be used in the calculations that
follow.


\subsection{A physical requirement for the relation between
foliation-dependent variables: the equivariance
condition}

As already mentioned, the histories approach allows the
realisation of a formalism that it is a hybrid of both the
covariant and the canonical  formalisms.

In \cite{Sav03} we discussed the loss of the spacelike character
of a foliation, under a change of the metric. Nevertheless, the
introduction of a metric-dependent foliation  solved the problem
\cite{Sav03}.

In addition to this, we need to address the partly
independent issue of the ${\rm Diff}(M)$-invariance---or,
rather, lack of ${\rm Diff}(M)$-invariance---of the
canonical variables. This is a problem even in the
standard Lagrangian treatment, because the solution of
the initial value problem requires the introduction of a
spacetime foliation. In the classical case, the question
concerning the dependence of physics on the choice of
foliation is easily resolved by specifying a unique
Lorentzian metric that solves the classical equations of
motion. However, in quantum theory the dependence of the
physical results on the choice of foliation is a major
issue.

In histories theory we define a representation of the
group of spacetime diffeomorphisms. The requirement of
the physical equivalence between  different choices of
time direction for the canonical theory, and the
requirement of  the spacetime character of the canonical
description, are satisfied by means of a simple
mathematical condition, namely  the {\em equivariance
condition\/} for the metric-dependent foliations.

To this end, we consider a particular class of
metric-dependent embeddings. We denote by $A(\cdot , g]$
any tensor field associated with the embedding, and which
is correspondingly a functional of the metric $g$. The
physical requirement is that the change of the tensor
field $A$ under a diffeomorphism transformation is
compensated by the change due to its functional
dependence on $g$.

Hence, if we consider a diffeomorphism transformation
$f$, and we denote its pull-back operation by $f^*$, the
equivariance condition is given by the expression
\begin{equation}
 (f^* A)(\cdot \,, g] = A(\cdot \,, f^* g].
\end{equation}
For an infinitesimal diffeomorphism transformation the
equivariance condition is
\begin{equation}
 {\cal L}_W A(X ; g] = \int\!\!d^4X^{\prime}\; \frac{\delta A(X;g]}
 {\delta g_{\mu\nu}(X^{\prime})} {\cal L}_W
 g_{\mu\nu}(X^{\prime}) \,.
\end{equation}

In the case of a function ${\cal E}:{\rm
LRiem}(M)\rightarrow {\rm Fol}(M)$, (where ${\rm
LRiem}(M)$ is the space of Lorentzian metrics on $M$, and
${\rm Fol}(M)$ is the space of foliations of $M$) we say
that ${\cal E}$ is an `equivariant foliation' if
\begin{equation}
                {\cal E}(f^*g)=f^{-1}\circ {\cal E}(g)
                \label{Def:EquivariantFoliation2}
\end{equation}
for all Lorentzian metrics $g$ and $f\in{\rm Diff}(M)$.

This concept has the following simple interpretation. Namely, we
consider the principle bundle ${\rm Diff}(M)\rightarrow {\rm
LRiem}(M)\rightarrow {\rm LRiem}(M)/{\rm Diff}(M)$, and then
define a left-action $\ell$ of ${\rm Diff}(M)$ on the space of
foliations ${\rm Fol}(M)$ by
\begin{equation}
        \ell_f({\cal F}):=f\circ {\cal F}
        \label{Def:ell}
\end{equation}
for all $f\in{\rm Diff}(M)$ and ${\cal F}\in {\rm Fol}(M)$. We use
this action to construct the associated bundle ${\rm
Fol}(M)\rightarrow {\rm LRiem}(M)\times_{{\rm Diff}(M)} {\rm
Fol}(M)\rightarrow {\rm LRiem}(M)/{\rm Diff}(M)$. Now, in general,
if $F\rightarrow P\times_GF\rightarrow X$ is a bundle associated
to a principle bundle $G\rightarrow P\rightarrow X$ (via a left
action of the group $G$ on $F$), the cross-sections of the
associated bundle are in one-to-one correspondence with functions
$\psi:P\rightarrow F$ satisfying the condition
\begin{equation}
        \psi(pg)=g^{-1}\psi(p)\label{XSection}
\end{equation}
for all $p\in P$ and $g\in G$. It follows, therefore,  from Eq.\
(\ref{Def:EquivariantFoliation2}), Eq.\ (\ref{Def:ell}) and Eq.\
(\ref{XSection}), that what we have called an `equivariant
foliation' is equivalent to a cross-section of the associated
bundle ${\rm Fol}(M)\rightarrow {\rm LRiem}(M)\times_{{\rm
Diff}(M)} {\rm Fol}(M)\rightarrow {\rm LRiem}(M)/{\rm Diff}(M)$
over the space ${\rm LRiem}(M)/{\rm Diff}(M)$ of ${\rm
Diff}(M)$-equivalence classes of Lorentzian metrics on $M$.

As we shall see, the use of equivariant foliations leads to a
significant result: the Hamiltonian constraints, the canonical
action functional, and the equations of motion on the reduced
state space are all invariant under the action of the group of
spacetime diffeomorphisms ${\rm Diff}(M)$.

\subsection{Relation between the invariance groups}

We have showed already that in histories theory there exist
representations of all three invariance groups of general
relativity. We now proceed to study the relations between them.

An immediate observation is that the group of spacetime
diffeomorphisms is a subgroup of the Bergmann-Komar group
${\cal{BK}}(M)$. Indeed, the generators of ${\rm
Diff}(M)$, given by the expression (\ref{Vw}), are a
special case of those of ${\cal{BK}}(M)$ where the vector
fields $W$ are {\em not\/} functionals of the metric
tensor $g$.

One of the deepest issues to be addressed in canonical gravity is
the relation of the algebra of constraints to the spacetime
diffeomorphisms group (and, therefore, to the enlarged group
${\cal{BK}}(M)$).

To this end, the commutator of the B-K (Bergmann-Komar) generators
with the canonical constraints can be written as
\begin{eqnarray}
 \{ U_W \,, \Phi(k) \} = \int \!\!d^4X d^4\! X' \left[ \left({\cal
 L}'_W \bar{E}^{\mu \nu}_{\rho \sigma}(X,X') \right. \right.
 \hspace{5cm}  \nonumber
 \\
 \left. \left. - \int \!\! d^4\!X''
 \frac{\delta \bar{E}^{\mu \nu}_{\rho \sigma}(X,X')}{\delta
 g_{\alpha \beta}(X'')} {\cal L}_W g_{\alpha \beta}(X'')\right)
\pi^{\rho \sigma}(X') n_{\mu}(X;g) k_{\nu}(X) \right.
  \hspace{2.3cm} \nonumber
  \\
\left. + \left( \bar{E}^{\mu \nu}_{\rho \sigma}(X,X') \pi^{\rho
\sigma}(X') k_{\nu}(X) \int \!\!d^4\! X'' \frac{\delta
n_{\mu}(X)}{\delta g_{\alpha \beta}(X'')} (- {\cal L}_W g_{\alpha
\beta}(X'') \right) \right. \hspace{1.7cm} \nonumber
\\
\left. + 2 \bar{E}^{\mu \nu}_{\rho \sigma}(X,X') n\!_{\mu}(\!X\!)
k_{\nu}(\!X\!)\! \int\!\! d^4\!X'' \frac{\delta
W^{\alpha}(X'')}{\delta g_{\rho \sigma}(X')} \nabla''_{\tau}
\pi^{\tau \beta}\!(X'') g_{\alpha \beta}(X'') \right]
\hspace{1.8cm} \label{Vwbk-phi}
\end{eqnarray}
\begin{eqnarray}
 \{ U_W \,, Q (\vec{L})\} = \int \!\!d^4\!X d^4\!X'
\left[ - \frac{\delta \bar{E}^{\mu \nu}_{\rho
\sigma}(X,X')}{\delta g_{\alpha \beta}(X'')} {\cal L}_W g_{\alpha
\beta}(X'') \pi^{\rho \sigma}(X') {\cal L}_W g_{\mu \nu}(X)
\right. \hspace{1.9cm}\nonumber
\\
\left. - \bar{E}^{\mu \nu}_{\rho \sigma}(\!X,X'\!) {\cal L}'\!_W
\pi^{\rho \sigma}(\!X'\!) {\cal L}\!_W g_{\mu \nu}(\!X\!) \right.
\left. \!\!{\cal L}\!_W g_{\mu \nu}(\!X\!) \pi^{\rho
\sigma}(\!X'\!) ({\cal L}\!_{{\delta_W}L} g_{\mu \nu}(\!X\!)\! -\!
{\cal L}\!_W {\cal L}\!_L g_{\mu \nu}(\!X\!) \right.
\hspace{1cm}\nonumber
\end{eqnarray}
\begin{equation}
\left. + 2 \bar{E}^{\mu \nu}_{\rho \sigma}(X,X') {\cal L}_W g_{\mu
\nu}(X)  \int d^4X'' \frac{\delta W^{\alpha}(X'')}{\delta g_{\rho
\sigma}(X')} \nabla''_{\tau} \pi^{\tau \beta}(X'') g_{\alpha
\beta}(X'') \right] \label{Vwbk-Hi}
\end{equation}
\begin{eqnarray}
 \{ U_W \,, {\cal{H}} (L)\} = \int d^4X  L(X)
 \nonumber \hspace{10.7cm}
 \\
\times \left[ \int \!d^4\!X' \!\frac{1}{2} \frac{\delta
\tilde{N}}{\delta  g_{\alpha \beta}(X')} (- {\cal
 L}_W g_{\alpha \beta})(X') G\!_{\mu \nu \rho \sigma}(X)(\bar{E}\pi)^{\mu \nu}(X)
(\bar{E}\pi)^{\rho \sigma}(X) \right. \nonumber \hspace{3.7cm}
 \\
\left. +\frac{\tilde{N}}{2}  \frac{\partial G\!_{\mu \nu \rho
\sigma}}{\partial h_{\kappa \lambda}}(\!X\!) \left(\delta^{\kappa
\lambda}_{\alpha \beta} - \!\int\! d^4\!X' \frac{\delta
(n^{\kappa} n^{\lambda})(\!X\!)}{\delta g_{\alpha \beta}(\!X'\!)}
{\cal L}_W g_{\alpha \beta}(\!X'\!) \right)(\bar{E}\pi)\!^{\mu
\nu}(\!X\!) (\bar{E}\pi)\!^{\rho \sigma}(\!X\!) \right.
 \nonumber \hspace{2.7cm}
\\
\left. - \int \!\!d^4\!X'  \tilde{N}(\!X\!) G_{\mu \nu \rho
\sigma}(\!X\!) \int \!d^4\!X'' \frac{\delta \bar{E}^{\mu
\nu}_{\kappa \lambda}(\!X,X'\!)}{\delta g_{\alpha \beta}(\!X''\!)}
{\cal L}_W g_{\alpha \beta}(\!X''\!) \pi^{\kappa \lambda}(\!X'\!)
(\bar{E}\pi)^{\rho \sigma}(\!X\!)\right. \nonumber \hspace{3.1cm}
 \\
\left. - d^4X' \tilde{N}(X) G_{\mu \nu \rho \sigma}(X)
\bar{E}^{\mu \nu}_{\kappa \lambda}(X,X') {\cal L}_W' \pi^{\kappa
\lambda}(X') (\bar{E}\pi)^{\rho \sigma}(X) \right. \nonumber
\hspace{5.5cm}
\\
\left. + \int \!d^4\!X' \frac{\delta \tilde{N}^{-1}}{\delta
g_{\alpha \beta}(X')}  {\cal L}_W g_{\alpha \beta}(X') {}^3R(X)
\right. \hspace{9.1cm} \nonumber
\\
 \left. - \tilde{N}^{-1}(X) {}^3R^{\alpha \beta}(X)
 \left( \delta^{\alpha \beta}_{\kappa
 \lambda}- \int d^4X' \frac{\delta ( n^{alpha}
 n^{\beta})(X)}{\delta g_{\kappa \lambda}(X')} {\cal L}_W g_{\kappa
 \lambda}(X')\right) \right.   \hspace{4.4cm} \nonumber
 \\
 \left. +2 \int\! d^4\!X' \tilde{N}(X) G_{\mu \nu \rho \sigma}(X)
 \bar{E}^{\mu \nu}_{\kappa \lambda}(X,X') \right. \hspace{9cm}\nonumber
 \end{eqnarray}
 \begin{equation}
 \times \left. \left(\int \!d^4\!X'' \frac{\delta W^{\tau}(X'')}
 {\delta g_{\kappa
 \lambda}(X')}\nabla''_{\alpha}\pi^{\alpha \beta}(X') g_{\beta
 \tau}(X'') \right) (\bar{E}\pi)^{\rho
 \sigma}(X) \right].  \hspace{2cm} \label{Vwbk-H}
\end{equation}

In these equations we have denoted by $\delta_W L$ the
total variation of $L$ under the action of the
infinitesimal  diffeomorphism generated by the vector
field $W$:
\begin{eqnarray}
\delta_W L^{\mu}(X) = ({\cal L}_W L)^{\mu}(X) - \int
d^4X' \frac{\delta L^{\mu}(X)}{\delta g_{\alpha
\beta}(X')} {\cal L}_W g_{\alpha \beta}(X') \,,
\end{eqnarray}
where $\tilde{N} := N / \sqrt{-g}$.

We note that we have considered the generator $Q(\vec{L})$ rather
than the supermomentum constraint, since the latter is a linear
combination of $Q(\vec{L})$ with the primary constraint $\Phi$.

The commutators of the ${\rm Diff}(M)$ generators with the
constraints emerge as a special case of the above equations. In
particular,
\begin{eqnarray}
 \{ V_W \,, \Phi(k) \} =  \Phi({\cal L}_Wk)+
\int \!\!d^4 \!X d^4\!X'  \delta_W [\bar{E}^{\mu \nu}_{\rho
\sigma}(\!X,X'\!) n_{\mu}(\!X\!)] \pi^{\rho \sigma}(\!X'\!)
k_n(\!X\!) \label{Vw-phi}
\end{eqnarray}
\begin{eqnarray}
 \{ V_W \,, Q (\vec{L})\} = Q(\delta_W \vec{L}) + \int \!\! d^4 \!X d^4\!X'
 \delta_W \bar{E}^{\mu \nu}_{\rho \sigma}(X,X') {\cal L}_L g_{\mu\
 \nu} \hspace{2.1cm}\label{Vw-Hi}
  \end{eqnarray}
\begin{eqnarray}
 \{ V_W \,, {\cal{H}} (L)\}=  {\cal H}({\cal L}_WL) \hspace{11cm}
 \nonumber
 \\
 + \!\!\int \!d^4\!X
 d^4\!X' d^4\!X'' L(\!X\!)\pi^{\mu \nu}(\!X'\!) \pi^{\kappa
 \lambda}(\!X''\!)\delta_W \left(\tilde{N}(\!X\!) G\!_{\mu \nu
 \rho \sigma}(\!X\!)
 \bar{E}^{\mu \nu}_{\kappa \lambda}(\!X,X'\!)\bar{E}^{\mu \nu}_{\kappa
 \lambda}(\!X,X'\!)\right) \hspace{2cm} \nonumber
 \end{eqnarray}
 \begin{equation}
+ \int d^4X L(X) \delta_W(\tilde{N}^{-1}{}^3R). \hspace{6.5cm}
 \label{Vw-H}
\end{equation}
We denote by $\delta_W$ total variation with respect to
the the vector field $W$. Then
\begin{equation}
\delta_W \bar{E}^{\mu \nu}_{\rho \sigma}(\!X,X'\!)= (\!{\cal L}_W
\!+ \!{\cal L}_W'\!) \bar{E}^{\mu \nu}_{\rho \sigma}(\!X,X'\!)  -
\!\int \!\!d^4\!X '' \frac{\delta \!\bar{E}^{\mu \nu}_{\rho
\sigma}(\!X,X'\!)}{\delta \!g_{\alpha \beta}(\!X''\!)} {\cal L}_W
g_{\alpha \beta}(\!X''\!),
\end{equation}
\begin{equation}
 \delta_W n_{\mu}(X) = {\cal L}_W n_{\mu} (X) - \int \!\! d^4\!X'
 \frac{\delta n_{\mu}(X)}{\delta g_{\alpha \beta}(X')}{\cal L}_W
 g_{\alpha \beta}(X'). \hspace{1.4cm}
\end{equation}
>From the above equations we conclude that the action of the
diffeomorphism group on the constraints amounts to the action of
the diffeomorphisms on the metric-dependent foliation. Hence, the
diffeomorphism group  generates transformations between the
reduced phase spaces that correspond to different metric-dependent
foliations.

Next, we impose the equivariance condition on the
foliation, which implements the physical principle that
histories canonical field variables related by
diffeomorphism transformations are physically equivalent.
One can show that the terms $\delta_W \bar{E}$ and
$\delta_W n$ vanish, and we get
\begin{eqnarray}
 \{ V_W \,, \Phi(k) \} &=&   \Phi({\cal L}_Wk) \label{equivVw-phi}  \\
 \{ V_W \,, Q(\vec{L})\} &=&  Q(\delta_W L) \label{equivVw-Hi}  \\
 \{ V_W \,, {\cal{H}} (L)\} &=&  {\cal H}({\cal L}_WL) \label{equivVw-H}.
\end{eqnarray}
Under the infinitesimal symplectic transformation
generated by $V_W$, the constraints transform  from
$\Phi(k), Q(\vec{L}), {\cal H}(L)$ to $\Phi(k'),
Q(\vec{L}'), {\cal H}(L')$, where
\begin{eqnarray}
k' = k + s {\cal L}_W k \\
\vec{L}' = \vec{L} + s \delta_W \vec{L} \\
L' = L + s {\cal L}_W L
\end{eqnarray}
In particular, if $\vec{L}^{\mu} n_{\mu} = 0 $, then
$(\delta_W \vec{L}^{\mu}) n_{\mu} = -  \vec{L}^{\mu}
\delta_Wn_{\mu} = 0$, where the vanishing of the last
term is due to the equivariance condition for $n_{\mu}$.
This implies that $\vec{L}'^{\mu} n_{\mu} =0$.

>From the above, we conclude that {\em the constraints of
canonical general relativity are ${\rm
Diff}(M)$-invariant\/}.

We mentioned in section 2.1.2 that Bergmann and Komar
claim that the Dirac algebra is a subalgebra of the
Bergmann-Komar algebra. However, this relation was not
established in \cite{BK} by comparing concrete
representations of these algebras. In fact, in our case
this relation does not hold, and the generators of the
Dirac algebra are {\em not\/} special cases of generators
of the B-K group. This is because the superhamiltonian is
quadratic in momentum, while the B-K group is linear in
momentum.

We should also note that the generators of the B-K group
do not commute with the constraints, hence the B-K group
is not realised on the reduced state space. This leads to
the question of whether there exists a different
representation of the B-K group which does commute with
the constraints (for equivariant foliations) and which
has the Dirac algebra as subalgebra. We leave this as an
open question.

\section{Reduced state space}
The study of the parameterised particle system has been
proved a helpful model\footnote{The parameterised
particle model is consider a good precursor for more
complicated parameterised systems, with of course general
relativity as the most complicated one.} for developing a
histories reduced state space algorithm for general
relativity.

We present first, very briefly, the histories reduced
state space algorithm, that was originally given in
\cite{SA00}.

\subsection{Histories treatment of constraints}

\paragraph*{Classical parameterised systems.}
Parameterised systems have a vanishing Hamiltonian $H = h(x , p)
$, when the constraints are imposed. Classically, two points of
the constraint surface $C$ correspond to the same physical state
if they are related by a transformation generated by the
constraint. The true degrees of freedom correspond to equivalence
classes of such points and are represented by points of the
reduced state space ${\Gamma}_{red}$.

An element of the reduced state space is {\em itself \/}
a solution to the classical equations of motion, and it
also corresponds to a possible configuration of the
physical system at an instant of time; hence the notion
of time is unclear, and it is not obvious how to recover
the notion of temporal ordering unless we choose an
arbitrary gauge-fixing condition.

In the histories approach to parameterised systems, the history
constraint surface $C_h$ is defined as the set of all smooth paths
from the real line to the constraint surface $C$. The history
Hamiltonian constraint is defined by $H_{\kappa} = \int \! dt\,
\kappa (t) h_t$, where $h_t := h(x_t, p_t)$ is first-class
constraint. For all values of the smearing function $\kappa (t)$,
the history Hamiltonian constraint $H_{\kappa}$ generates
canonical transformations on the history constraint surface $C_h$.
The history reduced state space $\Pi_{red}$ is then defined as the
set of all smooth paths on the canonical reduced state space
$\Gamma_{red}$, and it is identical to the space of orbits of
$H_{\kappa}$ on $C_h$.

In order for a function on the full state space, $ \Pi $,
to be a physical observable ({\em i.e.,\/} to be
projectable into a function on $ {\Pi}_{red}$), it is
necessary and sufficient that it commutes with the
constraints on the constraint surface.

Contrary to the canonical treatments of parameterised systems, the
classical equations of motion are explicitly realised on the
reduced state space $\Pi_{red}$. They are given by
\begin{equation}
\{ \tilde S , F \}\,(\gamma_{cl}) = \{ \tilde{V} , F
\}\,(\gamma_{cl}) = 0
\end{equation}
where $\tilde{S}$ and $\tilde{V}$ are respectively the action and
Liouville functions projected on $\Pi_{red}$. Both $\tilde{S}$ and
$\tilde{V}$ commute weakly with the Hamiltonian constraint
\cite{SA00}. Furthermore, the equations of motion on $\Pi_{red}$
remain invariant under time reparameterisations.

Hence, in the histories formalism, parameterised systems
have  an intrinsic time that does not disappear when we
enforce the constraints, either classically or quantum
mechanically.

\subsection{${\rm Diff}(M)$ invariance of the reduced state space}

We showed in section 3.4 that the generators $V_W$ of the
spacetime  diffeomorphisms group commute with the constraints, and
hence they are defined in the reduced state space $\Pi_{red}$.

The generator of time translations of the canonical theory is the
`Liouville' functional $V$,
\begin{equation}
 V:= \int \!\! dt \int \!\! d^3 \underline{x}  \left \{
{\tilde{\pi}}^{ij} (t,\underline{x})\,
{\dot{h}}_{ij}(t,\underline{x}) + {\tilde{p}}_i {\dot{N}}^i +
\tilde{p} \dot{N} \right \}.
\end{equation}

It can be easily shown that the Liouville function
commutes with the canonical constraints
\begin{eqnarray}
 \{ V , \Phi (k)\} &=& 0 , \\
 \{ V , {\cal{H}} (L)\} &=& 0 ,  \\
 \{ V , {\cal{H}} (\vec{L})\} &=& 0 .
\end{eqnarray}

The Liouville function can be written covariantly as
\begin{eqnarray}
V = \int d^4X  (\bar{E} \pi)_{\rho \sigma} \left[ h^{\mu \rho}
h^{\nu \sigma} {\cal L}_t g_{\mu \nu} + {\cal L}_t (n^{\rho}
n^{\sigma}) \right].
\end{eqnarray}
Here $t$ denotes the deformation vector $t^{\mu}$ associated to our foliation.

In this form, it is easy to show that the Liouville
function commutes with the diffeomorphisms:
\begin{eqnarray}
\{V, V_W \} = 0,
\end{eqnarray}
provided that the metric-dependent foliation satisfies
the equivariance condition.

The canonical action functional $S$ is defined as
\begin{eqnarray}
S &:=& \int \!\! dt\int \!\! d^3 \underline{x}  \left \{
{\tilde{\pi}}^{ij} (t,\underline{x})\,
{\dot{h}}_{ij}(t,\underline{x}) + {\tilde{p}}_i {\dot{N}}^i +
\tilde{p} \dot{N}- {\cal{H}} (N) - {\cal{H}} (\vec{N}) \right\}
\,,\label{Shis},
\end{eqnarray}
and can clearly be projected onto $\Pi_{red}$.

The classical equations of motion \textit{can} be explicitly
realised on the reduced state space $\Pi_{red}$. They are given by
\begin{equation}
\{ \tilde S , F \}\,(\gamma_{cl}) = \{ \tilde{V} , F
\}\,(\gamma_{cl}) = 0
\end{equation}
where $\tilde{S}$ and $\tilde{V}$ are respectively the action and
Liouville functions projected on $\Pi_{red}$.

The usual dynamical equations for the canonical fields $h_{ij}$
and $\pi^{ij}$ are equivalent to the history Poisson bracket
equations
\begin{eqnarray}
\{ S \,, h_{ij} (t,\underline{x}) \}\, (\gamma_{cl}) &=& 0
            \label{eqsSh} \\[2pt]
\{ S \,, \pi^{ij} (t,\underline{x}) \}\, (\gamma_{cl}) &=& 0
\label{eqsSpi}
\end{eqnarray}
where $S$ is defined in Eq.\ (\ref{Shis}). The path $\gamma_{cl}$
is a solution of the classical equations of motion, and therefore
corresponds to a spacetime metric that is a solution of the
Einstein equations.

Finally, let us note that the canonical action functional
$S$ is also diffeomorphic-invariant:
\begin{equation}
 \{ V_W , S \} = 0.
\end{equation}

The invariance of the action under diffeomorphisms is a
rather significant result, and it leads to the
conclusion, as one would have anticipated, that the
action functional and the equations of motion
(\ref{eqsSh}--\ref{eqsSpi}) are the `observables' of
general relativity theory, as has been indicated from the
Lagrangian treatment of the theory.

We showed that the Liouville function commutes with the
Hamiltonian constraints, hence it corresponds to a function
$\tilde{V}$ on the reduced state space. In fact, $\tilde{V}$
coincides with the projection of $S$ on $\Pi_{red}$.

The key point is that the elements of the reduced state space in
the histories formalism are also paths on the standard canonical
state space. As such, they preserve the notion of time, in the
sense that they are labelled by the external time parameter $t\in
\mathR$.

It is important to remember that the parameter with respect to
which the orbits of the constraints are defined, is not in any
sense identified with the physical time $t$. In particular, one
can distinguish the paths corresponding to the classical equations
of motion by the condition
\begin{equation}
 \{ F , \tilde{V}\}_{{\gamma}_{cl}} = 0 , \label{FV}
\end{equation}
where $F$ is a functional of the field variables, and
${{\gamma}_{cl}}$ is a solution to the equations of motion.

In standard canonical theory, the elements of the reduced
state space are all solutions to the classical equations
of motion. In histories canonical theory, however, an
element of the reduced state space is a solution to the
classical equations of motion only if it also satisfies
the condition Eq.\ (\ref{FV}). The reason for this is
that the histories reduced state space ${\Pi}_{red}$
contains a much larger number of paths (essentially all
paths on ${\Gamma}_{red}$ ). For this reason, histories
theory may naturally describe observables that commute
with the constraints but which are not solutions to the
classical equations of motion.

This last point should be particularly emphasised,
because of its possible corresponding quantum analogue.
We know that in quantum theory, paths may be realised
that {\em are not\/} solutions to the equations of
motion. My belief is that the histories formalism will
distinguish between instantaneous laws\footnote{A
thorough analysis on the connection between instantaneous
laws (Gauss' {\em theorema egregium}) and the dynamical
laws of general relativity is presented by Kuchar in
\cite{Kucinst}, where he deduces Einstein's equations
starting from the (instantaneous) geometric description
of gravity.} (namely constraints), and dynamical laws
(equations of motion).

Hence, it is possible to have a quantum theory for which the
instantaneous laws are satisfied, while the classical dynamical
laws are not. This distinction is present, for example, in the
history theory of the quantised electromagnetic field, where all
physical states satisfy the Gauss law exactly, however
electromagnetism field histories are possible which do not satisfy
the dynamical equations, i.e., Maxwell's equations. For
parameterised systems, this distinction is not possible within the
canonical formalism, nevertheless as we explained, it does arise
in the histories formalism.

The equations of motion (\ref{FV}) imply that physical observables
have constant values on the solutions to the classical equations
of motion. This need not be the case quantum mechanically, hence
quantum realised paths need not be characterised by `frozen'
values of their physical parameters.

\section{Some notes on quantisation}

It is interesting to examine the histories perspective on
quantisation, in the light of these new results.  We recall that
the canonical quantisation scheme is based upon the canonical
commutation relations and a search for their representations on a
Hilbert space. However, the canonical commutation relations come
originally from considering a {\em spacetime\/} metric, with the
associated requirement that the induced 3-metric is spacelike; and
one facet of the `problem of time' is to recover a spatio-temporal
picture from the purely spatial perspective of the strict
canonical formalism.

In the histories approach, this issue can be addressed fully. We
seek a representation of the {\em history commutation relations},
which are defined with reference to the whole of spacetime and not
just a 3-surface; in particular, these history variables include a
quantised Lorentzian spacetime metric. Finding a representation of
the history algebra on a Hilbert space $H$, makes plausible the
possibility of finding a representation the group ${\rm Diff}(M)$
on $H$, and possibly also the Bergmann-Komar group.

In particular, in the histories approach, we can manifestly
address a problem that is usually sidestepped in canonical
quantisation: namely how to compare the quantum schemes that come
from different choices of the `internal' time variable that must
be identified to recover a spacetime perspective. This is done by
the introduction of the metric-dependent foliations discussed
above, which makes explicit the relation between canonical and
covariant objects.

The next step is to write the representation of the constraint
generators, preferably in a way that the Dirac algebra is
preserved. As yet, the history procedure does not provide any
preferred strategy for doing this. However, we note that while one
may write a quantum history description for any canonical theory
\cite{Ana01} (characterised by the presence of a vacuum
state)\footnote{If $H$ is the Hilbert space of the canonical
theory, the corresponding history Hilbert space is a suitable
version of the "continuous tensor product" $\otimes_t H_t$}, there
exist possible representations of the history algebra, that do
{\em not} have any canonical analogue.

This is particularly relevant to the `loop quantum gravity'
approach to quantisation (for its basic features see \cite{loop},
and also \cite{Thiem} for a recent review). In the canonical
treatment, the basic algebra is defined with reference to objects
that have support on loops in the three-dimensional surface
$\Sigma$. In the history version, the relevant objects would be
defined on two-dimensional hypersurfaces within the spacetime $M$.

Alternatively, one could consider path variables corresponding to
the $SL(2,{\bf C})$ connection on the spacetime $M$, rather than
the $SU(2)$ one of the canonical theory. In that case the
representation theory of the history description would be very
different from the one of the canonical approach, mainly because
the $SL(2,{\bf C})$ group is non-compact.

In all cases, the mathematical structures of a quantisation based
on histories will conceivably be very different from those in the
canonical theory. For this reason, the history construction may
uncover substantially different properties from those that arise
in the existing approaches to loop quantum gravity. Given that the
history approach successfully addresses several key issues that
plague the canonical perspective, I believe it provides a
promising new approach to tackling the dynamical aspects of the
loop quantum gravity programme.

Once we have the constraint operators, we can look for the
physical Hilbert space. Technically, this is a problem of the same
degree of difficulty as the one that arises in the canonical
scheme. However, unless there exist some anomalies in the
algebraic relation of the spacetime diffeomorphism group to the
constraints, we expect the generators of the diffeomorphism group
to exist on the physical Hilbert space, together with the action
functional which generates the physical time translations. In this
way, the resulting theory will carry both the diffeomorphism
symmetry and have a well-defined notion of time-ordering, thus
solving another facet of the problem of time in canonical quantum
gravity. We shall discuss some simple systems of this type in
future work.

However, in order to realise the history quantisation scheme for
general relativity, we have to face again the problem of
constructing an operator representing the Hamiltonian constraint,
which is a formidable task. Or, at least, this would be so if we
followed the common wisdom for the quantisation of constrained
systems, namely a version of the Dirac approach, upon which our
history quantisation algorithm is also based. However, the history
theory has more versatility and may provide the concepts and
physical predictions of a full quantum theory without needing a
Hilbert space structure; for example, by exploiting the
geometrical objects on the classical phase space \cite{Ana03}.
This provides another potential avenue for quantisation, which may
sidestep the problems associated with the intricate construction
of a super-Hamiltonian operator.

\section{Conclusions}

Histories theory in general is characterised by two key
ingredients: the history group, and the existence of two distinct
generators of time transformations. A significant consequence of
this structure is the coexistence of a spacetime and canonical
description of a theory. In the case of gravity, this is reflected
in the existence of realisations of both the group of spacetime
diffeomorphisms, and the Dirac algebra of constraints.

In this paper, we have discussed the relation of these two
transformation groups. We focused on the physical equivalence of
solutions to the equations of motion associated with different
choices of foliation: specifically, if different descriptions are
to be equivalent, they should be related by spacetime
diffeomorphisms.

We showed how this physical requirement can be satisfied by the
introduction of the novel mathematical idea of equivariant
foliations. We showed that an immediate consequence is the
invariance under spacetime diffeomorphisms of the canonical
constraints, the reduced state space action functional, and the
equations of motion.

Furthermore, we discussed the enlarged symmetry group of spacetime
mappings that are functionals of the four-metric, originally
defined by Bergmann and Komar, and we showed that there also
exists a representation of this group in the histories theory.

Our results strongly suggest that a quantisation of general
relativity based on histories will provide a radically new
perspective on the problem of time that has so plagued the
canonical approach. In this sense, this paper provides a stepping
stone towards the construction of a quantum theory of gravity,
that is based on genuine spacetime objects.

\vspace{1cm}

\noindent{\large\bf Acknowledgements}

\noindent I would like to especially thank Charis Anastopoulos for
very helpful discussions throughout this work. I would like to
also thank Chris Isham for his help in the editing of this paper.
I gratefully acknowledge support from the EPSRC GR/R36572 grant.


\begin{thebibliography}{99}


\bibitem{Sav03} Ntina Savvidou.
\newblock General relativity histories theory I: The Spacetime Character of the Canonical
Description.
\newblock Work in preparation. May (2003).

\bibitem{BK} P. G. Bergmann and A. Komar.
\newblock  The coordinate group symmetries of general relativity.
\newblock  {\em Int. J. Th. Phys.}, Vol. 5, No. 1 (1972), pp.
15-28.



\bibitem{hole} J, Earman and J. Norton.
\newblock What price spacetime substantialism?
\newblock {\em Brit. Jour. Phil. Science} 38: 515, 1987;
\newblock J. Stachel.
\newblock Einstein's search for general covariance.
\newblock in "Einstein and the history of General Relativity: Vol
I",
\newblock edited by D. Howard and J. Stachel.
\newblock Birkh\"auser, Boston, 1989.



\bibitem{SavGR01} Ntina Savvidou. {\em General Relativity Histories Theory:
 Spacetime Diffeomorphisms and the Dirac Algebra of Constraints},
 Class. Quant. Grav. 18, 3611, 2001.

\bibitem{SA00} C. Anastopoulos and K. Savvidou.  \newblock  Histories quantisation of
parametrised systems: I. Development of a general algorithm.
\newblock {\em Class.\ Quant.\ Grav.} 17: 2463, (2000).


\bibitem{Kucinst} K. Kuchar.
\newblock The problem of time in canonical quantization.
\newblock  In {  A. Ashtekar \& J. Stachel, eds, `Conceptual Problems of
Quantum Gravity', Birkh$\ddot{a}$user, Boston, pp. 141-171.}


\bibitem{Sav99} K. Savvidou. \newblock The action operator in continuous time
histories. \newblock  {\em J. Math. Phys.}  40: 5657, (1999); \\
\newblock  K. Savvidou.
\newblock  Continuous Time in Consistent Histories. \newblock  {PhD Thesis}
in gr-qc/9912076, (1999).



\bibitem{Kuc91} K. Kuchar.
\newblock  Time and Interpretations of Quantum Gravity.
\newblock {\em Winnipeg 1991, Proceedings, General Relativity and Relativistic
Astrophysics}, 211


\bibitem{I92} C. J. Isham. \newblock   Canonical Quantum gravity and the Problem of
Time. \newblock In {GIFT Seminar 1992}:0157-288. gr-qc/9210011.



\bibitem{Sav01} K. Savvidou.  Poincar\'e invariance for continuous-time histories.
\newblock J. Math. Phys. 43, 3053, 2002.



\bibitem{Ana01} C. Anastopoulos.
\newblock Continuous-time Histories:
 Observables, Probabilities, Phase Space Structure and the Classical
 Limit.
 \newblock {\em J.Math.Phys.} 42: 3225, 2001.


\bibitem{loop} A. Ashtekar.
\newblock New Variables for Classical and
Quantum gravity.
\newblock  {\em Phys. Rev. Lett.} 57: 2244, 1986;
C. Rovelli and L. Smolin.
\newblock Loop Space representation of Quantum General Relativity.
\newblock  Nucl. Phys. B331, 80, 1990;
 A. Ashtekar, J. Lewandowski, D. Marolf, J. Mourao and T. Thiemann.
\newblock Quantization of Diffeomorphic Invariant Theories of Connections with Local Degrees of Freedom.
\newblock J. Math. Ph. 36, 6456, 1995;
C. Rovelli and L. Smolin.
\newblock Spin Networks and Quantum Gravity.
\newblock Phys. Rev. D52, 5743, 1995.

\bibitem{Thiem} T. Thiemann.
\newblock Introduction to Modern Canonical Quantum General
Relativity.
\newblock gr-qc/0110034.

\bibitem{Ana03} C. Anastopoulos.
\newblock Quantum processes on phase space.
\newblock {\em Ann. Phys.} 303: 275, 2003.






\end{thebibliography}
\end{document}